\documentclass[12pt]{article}

\usepackage{PRIMEarxiv}

\usepackage{setspace}

\usepackage[utf8]{inputenc} 
\usepackage[T1]{fontenc}    

\usepackage{hyperref}       
\usepackage{url}            
\usepackage{booktabs}       
\usepackage{amsfonts}       
\usepackage{nicefrac}       
\usepackage{microtype}      
\usepackage{lipsum}
\usepackage{graphicx}       
\graphicspath{{graphs/}}     
\usepackage{caption}
\usepackage{subcaption}
\usepackage[dvipsnames]{xcolor}
\usepackage{epsfig}
\usepackage{textcomp}

\usepackage{algorithm}
\usepackage{algorithmic}
\usepackage{dsfont}
\usepackage{natbib}
\usepackage{mathrsfs}
\usepackage{multirow, makecell}
\usepackage{tabularray}
\usepackage{soul}
\usepackage{float}
\usepackage{cite}

 \usepackage{amsmath}

\title{On-Chain Credit Risk Score in Decentralized Finance\\
}

\author{
  by \\
 \textbf{Rik Ghosh, Arka Datta, Sudipan Sinha, Vidhi Aggarwal, Rajdeep Sengupta} \\
  Chainrisk, UAE \\
    \AND
  Correspondence to: arka@chainrisk.cloud\\
}

\begin{document}

\maketitle



\begin{abstract} 

Decentralized Finance (DeFi), a financial ecosystem without centralized controlling organization, has introduced a new paradigm for lending and borrowing. However, its capital efficiency remains constrained by the inability to effectively assess the risk associated with each user/wallet. This paper introduces the 'On-Chain Credit Risk Score (OCCR Score) in DeFi', a probabilistic measure designed to quantify the credit risk associated with a wallet. By analyzing historical real-time on-chain activity as well as predictive scenarios, the OCCR Score may enable DeFi lending protocols to dynamically adjust Loan-to-Value (LTV) ratios and Liquidation Thresholds (LT) based on the risk profile of a wallet. Unlike existing wallet risk scoring models, which rely on heuristic-based evaluations, the OCCR Score offers a more objective and probabilistic approach, aligning closer to traditional credit risk assessment methodologies. This framework can further enhance DeFi's capital efficiency by incentivizing responsible borrowing behavior and optimizing risk-adjusted returns for lenders.

\end{abstract}

\pagebreak 

\section{Introduction}

The ongoing digital revolution is driving a profound transformation in the financial sector. A key development in this evolution is decentralized finance (DeFi), which uses blockchain technology to facilitate open and permissionless financial services \citep{sahu2024secure}. DeFi signifies a paradigm shift in financial systems that reshapes the future of global finance. DeFi provides financial services without the intervention of any centralized intermediaries. They operate mainly as automated protocols on a blockchain \citep{doerr2021defi}. Although much more new in concept, DeFi is a fast-growing market for quick and safe interaction between lenders and borrowers. It is one of the best ways to seamlessly perform verifiable cross-border transactions in a much faster way compared to Traditional Finance. Numerous avenues can be explored to make DeFi more robust and capital-efficient. However, for such explorations, we need to understand the creditworthiness of each user. It should be noted that `user' means the specific wallet interacting with the DeFi ecosystem, and in this article, user and wallet are used interchangeably. 

Unlike conventional credit scoring, which depends on centralized credit bureaus and financial history of an user, DeFi credit scoring leverages on-chain data and smart contract interactions to evaluate an user's borrowing and repayment behavior. Credit risk scoring is a systematic process used to assess the risk involved in a user's creditworthiness by analyzing their on-chain transaction history, repayment behavior, outstanding liabilities, and other relevant economic indicators. This analytical approach enables financial institutions to quantify the probability of default, facilitating data-driven risk assessment and informed lending decisions \citep{mogherole}. An `On-Chain Credit Risk Score (OCCR Score)' for a wallet might be an answer to quantifying the credit risk of the particular wallet in DeFi ecosystem. Through the OCCR Score of a wallet, we have estimated the probability that the particular wallet may face liquidation when any borrow position is opened. 
Throughout the paper, we have used borrow positions and loans interchangeably.


There are different ways to give credit scores to users in the current TradFi landscape, but this is not very prevalent in DeFi. However, there have been previous attempts to measure the creditworthiness of a wallet, including those by Cred \citep{credprotocol}, Credit Data Alliance (CreDA) \citep{creda}, Credit scoring of Aave accounts \citep{wolf2022scoring} and Levon \citep{blockanalitica2024}. The main limitation of these existing credit scoring models is the absence of a comprehensive framework that integrates historical, current, and future predictive scenarios. The Cred Protocol scoring model combines financial metrics such as loan history, account composition, account health, and new credit with qualitative factors such as ecosystem participation and attributes related to trust and transparency, which are described in a very vague way \citep{packin2024decentralized}.  The CreDA credit scoring model relies heavily on social activity data, which could obscure more concrete financial indicators such as asset holdings, lending and borrowing behavior, and off-chain data \citep{creda}. This emphasis on decentralized social engagement raises concerns about the model's ability to provide a reliable and objective assessment of creditworthiness \citep{packin2024decentralized}. In addition, most of the existing scores take natural numbers on a variable scale. This scaling cannot be directly associated with a probability, which might make them subjective in nature. So, we have tried to develop a score that can bridge the three scenarios of historical, current, and predictive future and is equivalent to the probability of default for that particular wallet.



Section \ref{sec:OCCR Score} explains the overall formulation of the OCCR Score and of each subscore of the OCCR Score. A simulation study has been conducted using synthetically generated data in Section \ref{sec:simul}. Section \ref{sec:dynamic_ltv} discusses the usage of the OCCR Score to make the LTV/LF dynamic for each of the wallets. In addition, in the Appendix Section \ref{sec:appendix}, the different statistical properties of an estimate \citep{fisher1925theory} are derived in detail. By estimate, we mean the subscores of the OCCR Score.

\section{Framework}

In this section, we will discuss the overall framework of the OCCR Score and the notational explanations that will help us to develop the OCCR Score for the different wallets. The OCCR score, which is a probability of default of a particular wallet, has a range of $(0,1)$ and is a quantitative measure of the risk of default of the wallet's credit. The risk associated with a wallet is its unreliability of repayment when a borrow position is opened. Thus, to understand the OCCR Score of the wallet, we need to look at both the historical behavior of the wallet, the present dynamics of the wallet such as the current risk subscore, the credit utilization, and other factors, which are explained in detail in Section \ref{sec:OCCR Score}. Here, we go through the terms that have been used extensively in later Sections. $L_{i,j}$ denote the loan amount taken by the $i^{th}$ wallet corresponding to the $j^{th}$ loan/position. This random variable $L_{i,j}$ can take any positive real value. Since there might be single and multiple assets that can be provided as collaterals in borrowing, we need to associate different scores to each of the assets depending on the asset's riskiness. $r_{i,j}$ is used to denote the risk associated with each asset. Since this measure is of relative nature, it will have a range of $\left[0,1\right]$. Similarly to the loan amount, we also define $C_{i,j,k}$ as the amount provided by the $i^{th}$ wallet to maintain the position $j^{th}$ when paid on the $k^{th}$ asset. We denote the total current holding of a particular $i^{th}$ wallet by $H_{i}$. $T_{i,j}$ denotes the amount transacted by the $i^{th}$ wallet in the $j^{th}$ transaction. Similarly to the loan amount, all random variables $C_{i,j,k}$, $H_{i}$, and $T_{i,j}$ can take any real positive value.

\section{On-Chain Credit Risk (OCCR) Score} \label{sec:OCCR Score}

\subsection{Historical Credit Risk subscore $(\hat{s}_{h_i})$}

In this section, we will dive into analyzing the historical data for each wallet $i$. $X_{i,j}$ is a dichotomous random variable, defined as

\[
X_{i,j} =
\begin{cases}
  1 & \text{if loan/position is liquidated} \\
  0 & \text{if loan/position is repayed.}
\end{cases}
\]

It should be noted that $P(X_{i,j}) = s_{h_i}$. We need to estimate the parameter $s_{h_i}$. Thus, we observe the ratio estimate,

\begin{eqnarray}
  \hat{s}_{h_i} = \frac{\sum\limits_{j}w_{i,j}X_{i,j}}{\sum\limits_{j}w_{i,j}},
\end{eqnarray}

where $w_{i,j} = L_{i,j} \times \left(1-r_{i,j}\right) \times p_{i,j} \times t_{i,j}$. 

The combined riskiness of all collateral assets is calculated by $r_{i,j} = \frac{\sum \limits_{k} C_{i,j,k} \left(\frac{\sigma_{C_{i,j,k}}}{\sigma_{Max}}\right)}{\sum \limits_{k} C_{i,j,k}},$ for $C_{i,j,k}$ are the collaterals provided for the $j^{th}$ loan. Here, by `Maximum Volatility $(\sigma_{Max})$' we mean the maximum observed volatility among the collateral assets under consideration, while `Asset Volatility $(\sigma_{Asset})$' is the volatility of the corresponding collateral asset $(k)$ provided for the $j^{th}$ loan. 
$p_{i,j}$ denotes the proportion of the liquidated collateral asset (loan amount) as this proportion might vary depending on the particular DeFi protocol. The recency of the loan is represented by $t_{i,j}$. $t_{i,j}$ will be of the form of,

\begin{eqnarray}
  t_{i,j} = \frac{1}{1+e^{-\left(dt_{i,j}-k\right)}},
\end{eqnarray}
where $dt_{i,j}$ is the corresponding month of particular date when the loan position was commenced, and $k$ is such that the $t_{i,j}$ takes the value of $0.5$ for the month which falls in the middle of the whole period.

\subsection{Current Credit Risk subscore ($\hat{s}_{c_i}$)}
This section deals with the current (open) positions associated with the $i^{th}$ wallet. To understand the risk associated with the wallet, we need to understand the possibility of the particular wallet not being able to repay all the outstanding (open) loans if any unprecedented situation arises. Thus, to understand this, we observe the Liquidation at Risk (LaR) value for that particular wallet \citep{perez2021liquidations}. It is to be noted that if the total LaR observed across all loans/positions does not exceed the current holding of the wallet, then the wallet is safe and credible to get further loans. Now, to observe LaR for a particular position, we need to simulate the price path of different assets. Thus, we simulate price paths in sets of a certain number, say 2000, and continue to do so until the difference of the variance of LaR values are convergent. The convergence check is performed using the condition $|\sigma_{LaR}^{t+1}-\sigma_{LaR}^{t+1} \leq \epsilon|$. Let us define the random variable $Z_{i,j}$, which is a dichotomous variable. $Z_{i,j}$ takes the value $1$, if the total liquidations at risk ($LaR_{total}$) exceeds or equals to the current holding of the $i^{th}$ wallet, otherwise it is 0. Thus,

\[
Z_{i,j} =
\begin{cases}
  1 & \text{if} LaR_{total} \geq  H_{i} \\
  0 & \text{if} LaR_{total} <  H_{i}.
\end{cases}
\]
where $H_{i}$ is the current holding for the $i^{th}$ wallet, and $E(Z_{i,j})=s_{c_i}$.

The estimate of $s_{c_i}$ is given by,

\begin{eqnarray}
  \hat{s}_{c_i} = \frac{\sum\limits_{j=1}^{m} Z_{i,j}}{m},
\end{eqnarray}

, $m$ is the total number of times the price path has been simulated.

\subsection{Credit Utilization ($\hat{s}_{cu_{i}}$)}

In this section, we will explain the dependence of the subscore on the utilization of the available credit limit. We define the subscore for credit utilization, taking three different components into account, namely $C_{i,j}$ is the collateral asset (in $\$$) given during the opening of the position $j$, $LTV_{i,j}$ is the Loan-to-Value ratio prevalent at that exact time and $L_{i,j}$ as the loan amount taken at the $j^{th}$ position. 

The estimate of $s_{cu_i}$ is given by,

\begin{eqnarray}
    \hat{s}_{cu_i} = \frac{\sum_j\left(1-\left(\frac{L_{i,j}}{C_{i,j}\times LTV_{i,j}}\right) \right) \times L_{i,j} }{\sum\limits_{j}L_{i,j}}.
\end{eqnarray}

\subsection{On-Chain Transaction $(\hat{s}_{ct_i})$}

Here, we will try to understand the on-chain transaction of a particular wallet. We will mainly focus on the number and the size of the transaction, with a larger weight given to recent transactions. The on-chain transaction subscore associated with the OCCR Score is given by

\begin{eqnarray}
    \hat{s}_{ct_i} = \frac{\sum\limits_{l}T_{i,l}S_{i,l}t_{i,l}}{\sum\limits_{l}T_{i,l}},
\end{eqnarray}
where $T_{i,l}$ is the $l^{th}$ transaction amount of the $i^{th}$ wallet. If the transaction is credited to the wallet $i$, $S_{i,l}$ is $+1$, otherwise, the variable takes a value of $-1$.

\subsection{New Credit $(\hat{s}_{nc_i})$}

In this section, we will look at the risk associated with the wallet for taking recent loans in bulk (multiple times within a particular time span). We will be using cluster analysis to study the pattern of the wallet in taking multiple loans in past, and compare that with recent loans opened. Depending on the multiple loan positions opened compared to the earlier cases, a negative point will be assigned to the wallet, given by $\hat{s}_{nc_i}$.

We denote the loan amount for the $j^{th}$ loan of the $i^{th}$ wallet as $L_{i,j}$, and the corresponding date when the loan was taken as $D_{i,j}$. Now, the shortest interval of the $j^{th}$ loan compared to its two consecutive loans is denoted by $\Delta D_{i,j}$, where $\Delta D_{i,j}=min(\left(D_{i,j}-D_{i,j-1}\right),\left(D_{i,j+1}-D_{i,j}\right))$. Let the mean value of all the loan amounts taken by the wallet in the last month (tentative for now) be denoted by $\mu_{L_i}$, while the mean of the intervals be denoted by $\mu_{\Delta D_{i}}$. The total number of loans taken by the $i^{th}$ wallet in the last month is indicated by $n$. Let $Y_{i,j}$ be a dichotomous random variable such that it is given by

\[
Y_{i,j} =
\begin{cases}
  1 & \text{if } L_{i,j} \geq \mu_{L_i} \text{ and } \Delta D_{i,j} \leq \mu_{\Delta D_{i}}  \\
  0 & \text{if otherwise,}
\end{cases}
\]

with $P(L_{i,j} \geq \mu_{L_i} \text{ and } \Delta D_{i,j} \leq \mu_{\Delta D_{i}}) = P(L_{i,j} \geq \mu_{L_i}) \times P(\Delta D_{i,j} \leq \mu_{\Delta D_{i}}) = s_{nc_{i}}$.

Now, the new credit risk subscore is given by,
\begin{eqnarray}
    \hat{s}_{nc_i} = \frac{\sum\limits_{j=1}^{n} Y_{i,j}}{n}
\end{eqnarray}

\subsection{OCCR Score}

The OCCR Score for the particular wallet is found using a weighted average of all the above credit risk subscores obtained. The weight associated to each of the subscores is $0.35,0.25,0.15,0.15,0.10$ to $\hat{s}_{h_i},\hat{s}_{c_i},\hat{s}_{cu_i},\hat{s}_{ct_i}$ and $\hat{s}_{nc_i}$ respectively. Thus, the OCCR Score is obtained as



\begin{eqnarray}
    \text{OCCR Score} = 0.35\times \hat{s}_{h_i}+0.25\times \hat{s}_{c_i}+0.15\times (1-\hat{s}_{cu_i})-0.15\times \hat{s}_{ct_i}+0.10\times \hat{s}_{nc_i}
\end{eqnarray}

\section{Simulation} \label{sec:simul}

Any theoretical construe needs to be backed by a corresponding simulation study. In this section, we have tried to synthetically generate data for different wallets and understand the theoretical results claimed in Section \ref{sec:OCCR Score}. For the simulation study, we have taken different parameter values to generate transactions and borrowing positions for a wallet. The parameter values include the loan amount, the collateral amount, the timestamp, the asset provided as collateral, and others.

In our simulation study for the on-chain transaction subscore $(\hat{s}_{ct_i})$, we generated synthetic data for different wallets by varying key parameters such as the probability of credit transactions $p$. This probability is a measure of the probability that the wallet will have an amount credited to the wallet. Usually, transaction amounts are heavy-tailed in distribution. Thus, we assume that the transaction amount follows a Pareto distribution with shape parameter $\alpha$ and scale $x_{min}$. For each transaction, we randomly generated the amount of the transaction using a Pareto distribution (with the specified $\alpha$ and $x_{min}$) \citep{arnold2014pareto}, assigned a sign based on whether it was a credit or debit (which is also generated randomly from the Bernoulli distribution with a success probability of $p$). We weighted it by a recency score drawn uniformly from $[0,1]$. We then calculated the on-chain transaction subscore by taking the weighted sum of these transactions and normalizing it by the total transaction amount. For each wallet, we assumed a total of $60,000$ transactions. In addition, the process was repeated for a total of $5000$ times and the results obtained are tabulated in Table \ref{tab:on_chain_score}.

\begin{table}[hbt!]
\caption{Estimated On-Chain Transaction subscore ($\hat{s}_{ct_i}$) based on 5000 simulations along with corresponding Sample Standard Error (SSE), Asymptotic Standard Error (ASE), Coverage Probability (CP).}
\vspace{0.5cm}

\resizebox{\columnwidth}{!}{%
\renewcommand{\arraystretch}{2}%
 \begin{tabular}{|c|c|c|c|c|c|}
\hline
No. & $(p,\alpha,x_{min})$ & $\hat{s}_{ct_i} (s_{ct_i})$ & ASE & SSE & $\widehat{CP}$\\ \hline


1 & (0.60, 2.10, 300) & 0.0997 (0.10) & 0.000031 & 0.000019 & 0.985\\ \hline

2 & (0.35, 2.25, 300) & -0.1496 (-0.15) & 0.000014 & 0.000013 & 0.957\\ \hline
3 & (0.80, 2.10, 320) & 0.2991 (0.30) & 0.000023 & 0.000014 & 0.984\\ \hline
4 & (0.68, 2.60, 110) & 0.1796 (0.18) & 0.000008 & 0.000009 & 0.946\\ \hline
5 & (0.42, 2.06, 108) & -0.0798 (-0.08) & 0.000049 & 0.000021 & 0.992\\ \hline

\end{tabular}
\label{tab:on_chain_score}%
}

\end{table}

In Table \ref{tab:on_chain_score}, we have presented five different scenarios assuming different parameter values, which distinguish the specific wallet. In reality, we can observe that the borrowing or transaction patterns change from wallet to wallet. To mimic the different wallets, we have taken different values of the parameters such that all scenarios encompass both negative and positive values for on-chain transaction subscore. In all rows of Table \ref{tab:on_chain_score}, we can see that the theoretical values are nearly equal to the estimated ones. Also, in all rows, the coverage probability is near $0.95$ or higher, which means that the confidence interval constructed using the estimated values includes the theoretical mean in all the scenarios. Thus, it can be said that the estimate is both unbiased and reliable \citep{voinov2012unbiased}. Also, since the ASE and SSE values are very close to each other and also very low, it implies that the estimator is also more consistent.

\begin{table}[hbt!]
\caption{Estimated Credit Utilization subscore ($\hat{s}_{cu_i}$) based on 5000 simulations along with corresponding Sample Standard Error (SSE), Asymptotic Standard Error (ASE), Coverage Probability (CP).}

\vspace{0.5cm}
\resizebox{\columnwidth}{!}{%
\renewcommand{\arraystretch}{2}%
 \begin{tabular}{|c|c|c|c|c|c|}
\hline
No. & $(\alpha,x_{min},l_{min},l_{max})$ & $\hat{s}_{ct_i} (s_{ct_i})$ & ASE & SSE & $\widehat{CP}$\\ \hline


        1 & (2.10, 300, 0.50, 0.90)  & 0.333333 (0.333344) & 0.0000059  & 0.0000035  & 0.986  \\ 
        \hline
        2 & (2.30, 210, 0.64, 0.92) & 0.333359 (0.333338) & 0.0000025 & 0.0000022 & 0.965 \\ 
        \hline
        3 & (2.80, 50, 0.62, 0.84)  & 0.333341 (0.333336) & 0.0000014 & 0.0000014 & 0.952 \\ 
        \hline
        4 & (2.45, 680, 0.46, 0.74) & 0.333368 (0.333337) & 0.0000019 & 0.0000019 & 0.957 \\ 
        \hline
        5 & (2.45, 680, 0.46, 0.94) & 0.333351 (0.333337) & 0.0000020 & 0.0000018 & 0.958 \\ 
        \hline

\end{tabular}
\label{tab:credit_util}%
}

\end{table}

Table \ref{tab:credit_util} elaborates on the simulated results and compares those results with the theoretically obtained results for the credit utilization subscore $(\hat{s}_{cu_i})$. Here, we also observe that the theoretical and the simulated estimator values are nearly the same, with pretty low ASE and SSE values. Thus, we can surely say that the credit utilization subscore estimator is also unbiased, reliable, and consistent. Simulation studies for other subscores can be done along similar lines as above.

\section{Dynamic LTV Adjustment} \label{sec:dynamic_ltv}
Dynamic LTV adjustment means that the LTV ratio offered to a borrower changes based on their OCCR Score, which quantifies their creditworthiness. A wallet with a lower OCCR Score (indicating a good repayment history and low default risk) could be rewarded with a higher LTV ratio (more borrowing power against their collateral), while a wallet with a high OCCR Score would be offered the LTV ratio prevalent in the market at that time.

\subsection{Stochastic Modeling of Dynamic LTV}
Dynamic adjustment of LTV can be modeled as a stochastic process, which means that the LTV ratio is not fixed but changes over time, influenced by the borrower’s OCCR score. A stochastic process accounts for randomness and uncertainty in how factors such as market conditions or borrower behavior evolve, making the model adaptable to real-world scenarios.

\subsubsection{Basic Model Structure}
Let’s define a time-dependent LTV ratio, \( LTV(t) \), which is adjusted according to the borrower's OCCR score at time \( t \). We can express this as follows.

\begin{equation}
    LTV(t) = LTV_{\text{fixed}} - min(f(OCC_{\text{Score\_t}}),0)
\end{equation}

Where:
\begin{itemize}
    \item \( LTV(t) \) is the LTV ratio at time \( t \),
    \item \( LTV_{\text{fixed}} \) is the base or market-determined LTV ratio that would be applied in the absence of any OCCR score,
    \item \( f(OCC\_Score_t) \) is an adjustment function that modifies the LTV based on the OCCR Score at time \( t \).
\end{itemize}

\subsubsection{Adjustment Function \( f(OCCR\_Score_t) \)}
The adjustment function, \( f(OCCR\_Score_t) \), increases or decreases the LTV ratio depending on the borrower’s risk profile at any given time. It could take several forms, such as a linear or non-linear relationship between the OCCR score and the LTV ratio. For example:

\[
f(OCCR\_Score_t) = \alpha \cdot (OCCR\_Score_t - OCCR_{\text{avg}})
\]

Where:
\begin{itemize}
    \item \( \alpha \) is a scaling parameter that controls the sensitivity of LTV adjustments based on the OCCR Score,
Use past data to backtest how different \( \alpha \) values would have impacted loan performance and liquidation risks.
    
    \item \( OCCR\_Score_t \) is the OCCR score at time \( t \),
    \item \( OCCR_{\text{avg}} \) is the average OCCR score in the market.
\end{itemize}

\section{Conclusion}

In this paper, we have performed a detailed analysis of historical credit risk, current credit risk, new credit, credit utilization, and on-chain transaction subscores, providing valuable insight into their expectations, variances, and consistency. Under the assumptions outlined, each estimator is consistent in estimating the respective credit risk score for the wallet. In Section \ref{sec:simul} and Appendix \ref{sec:appendix}, we have tried to establish the unbiased \citep{voinov2012unbiased} nature of all subscores, using simulation and theoretical studies, respectively. Collectively, these estimators offer a comprehensive approach to assessing wallet risk across different time frames and transaction types, facilitating more accurate and reliable credit risk assessments. In Section \ref{sec:simul}, we have synthetically generated the transaction data for different wallets to emulate the real-life scenario and compared the simulated results with the theoretical results. For two of the subscores, it was observed that both results matched a high coverage probability score. Thus, it can be claimed that the estimators are reliable for practical applications.

The `On-Chain Credit Risk Score (OCCR Score)’ of wallets will help lending borrowing protocols and other DeFi institutes to understand the risk involved in allowing a wallet to open borrow position and thus may change the Loan-to-Value (LTV) ratio and subsequently the Liquidation Threshold (LT) if required. Through the OCCR score, we are trying to tailor the LT/LTV for particular wallets, hence enabling 'walletized finance'. If a lower `OCCR Scoring' is associated with a wallet, DeFi institutions may be incentivized to offer them borrow positions at a higher LT/LTV ratio than observed in the market, while for wallets maintaining a higher `OCCR Scoring' value, they may decide on keeping the LT/LTV ratio the same as the one prevalent in the market. This will encourage wallets to maintain a lower credit risk score to get loans at a much better (higher) LT/LTV ratio value than what is prevalent in the market. This might help the DeFi market to be more capital-efficient while maintaining a less risky approach. It should be noted that a wallet that enters the DeFi ecosystem for the very first time will receive a mean OCCR Score since that wallet has yet to make its first on-chain transaction.

\pagebreak

\section{Appendix} \label{sec:appendix}
\subsection{Expectation, Variance and Consistency of Historical Sub Score }

In this section, we derive the approximate expectation and variance of the historical subscore:
\[
\hat{s}_{h_i} = \frac{N_i}{D_i} = \frac{\sum_{j=1}^n w_{i,j} X_{i,j}}{\sum_{j=1}^n w_{i,j}},
\]
where for each loan (or position) \(j\):
\begin{itemize}
    \item \(X_{i,j}\) is a Bernoulli variable with 
    \[
    P(X_{i,j}=1)=s_{h_i},\quad P(X_{i,j}=0)=1-s_{h_i} \quad \Longrightarrow \quad \mathbb{E}[X_{i,j}]=s_{h_i}.
    \]
    \item The weight is given by
    \[
    w_{i,j}= L_{i,j} \, (1-r_{i,j}) \, p \, t_{i,j},
    \]
    with:
    \begin{itemize}
        \item {Loan Amount:}  
        \[
        L_{i,j}\mid (\text{ltv}, \text{collateral}) \sim \operatorname{Uniform}(0,\, \text{ltv}\times\text{collateral}),
        \]
        then
        \[
        \mathbb{E}[L_{i,j}\mid \text{ltv},\text{collateral}]=\frac{\text{ltv}\times\text{collateral}}{2}.
        \]
        If 
        \[
        \text{ltv}\sim\operatorname{Uniform}(l_{\min},l_{\max})\quad\text{and}\quad \mathbb{E}[\text{ltv}] = \frac{l_{\min}+l_{\max}}{2},
        \]
        and if the collateral is Pareto distributed with parameters \(\alpha\) and scale \(m\), so that
        \[
        \mathbb{E}[\text{collateral}] = \frac{\alpha\,m}{\alpha-1},
        \]
        then, by independence,
        \[
        \boxed{\mathbb{E}[L_{i,j}]=\frac{(l_{\min}+l_{\max})\,\alpha\,m}{4(\alpha-1)}.}
        \]
        Similarly, using \(\mathbb{E}[X^2]=\frac{a^2}{3}\) for a \(\operatorname{Uniform}(0,a)\) variable,
        \[
        \boxed{\mathbb{E}[L_{i,j}^2]=\frac{(l_{\min}^2+l_{\min}l_{\max}+l_{\max}^2)\,\alpha\,m^2}{9(\alpha-2)}.}
        \]
        
        \item {Risk Factor:} If \(r_{i,j}\sim\operatorname{Uniform}(0,1)\), then
        \[
        \mathbb{E}[1-r_{i,j}]=\frac{1}{2},\quad \mathbb{E}[(1-r_{i,j})^2]=\frac{1}{3}.
        \]
        
        \item {Recency:} Assuming \(t_{i,j}\sim\operatorname{Uniform}(0,1)\),
        \[
        \mathbb{E}[t_{i,j}]=\frac{1}{2},\quad \mathbb{E}[t_{i,j}^2]=\frac{1}{3}.
        \]
        
        \item {Liquidation Probability:} Here, \(p\) is a constant.
    \end{itemize}
    
    Under the assumption of independence of the components, the first and second moments of the weight are:
    \[
    \mathbb{E}[w_{i,j}] = \mathbb{E}[L_{i,j}]\cdot\frac{1}{2}\cdot p \cdot\frac{1}{2} = \frac{p\,\mathbb{E}[L_{i,j}]}{4},
    \]
    and
    \[
    \mathbb{E}[w_{i,j}^2] = \mathbb{E}[L_{i,j}^2]\cdot\frac{1}{3}\cdot p^2 \cdot\frac{1}{3} = \frac{p^2\,\mathbb{E}[L_{i,j}^2]}{9}.
    \]
\end{itemize}

\subsubsection{Expectation of the Historical Credit Risk Score}
Define
\[
N_i=\sum_{j=1}^n w_{i,j}X_{i,j} \quad \text{and} \quad D_i=\sum_{j=1}^n w_{i,j}.
\]
Since \(w_{i,j}\) and \(X_{i,j}\) are independent,
\[
\mathbb{E}[w_{i,j}X_{i,j}] = \mathbb{E}[w_{i,j}]\,\mathbb{E}[X_{i,j}] = s_{h_i}\,\mathbb{E}[w_{i,j}],
\]
so that
\[
\mathbb{E}[N_i] = s_{h_i} \sum_{j=1}^n \mathbb{E}[w_{i,j}],\quad \mathbb{E}[D_i] = \sum_{j=1}^n \mathbb{E}[w_{i,j}].
\]
Using the approximation
\[
\mathbb{E}\left[\frac{1}{D_i}\right] \approx \frac{1}{\mathbb{E}[D_i]} + \frac{\operatorname{Var}(D_i)}{\bigl(\mathbb{E}[D_i]\bigr)^3},
\]
we have
\[
\begin{aligned}
\mathbb{E}[\hat{s}_{h_i}] &\approx \mathbb{E}[N_i] \cdot \mathbb{E}\left[\frac{1}{D_i}\right] \\
&\approx \left(s_{h_i}\sum_{j=1}^n \mathbb{E}[w_{i,j}]\right)
\left(\frac{1}{\sum_{j=1}^n \mathbb{E}[w_{i,j}]} + \frac{\operatorname{Var}(D_i)}{\left(\sum_{j=1}^n \mathbb{E}[w_{i,j}]\right)^3}\right)\\[1mm]
&= s_{h_i} \left(1 + \frac{\operatorname{Var}(D_i)}{\left(\sum_{j=1}^n \mathbb{E}[w_{i,j}]\right)^2}\right).
\end{aligned}
\]
In the special case where all loans are identically distributed (denoting \(\mu_w = \mathbb{E}[w_{i,j}]\) and \(\mu_{w^2} = \mathbb{E}[w_{i,j}^2]\)), we have
\[
\mathbb{E}[D_i] = n\,\mu_w,\quad \operatorname{Var}(D_i) = n\Bigl(\mu_{w^2}-\mu_w^2\Bigr),
\]
so that
\[
\boxed{
\mathbb{E}[\hat{s}_{h_i}] \approx s_{h_i}\left( 1 + \frac{\mu_{w^2}-\mu_w^2}{n\,\mu_w^2}\right).
}
\]
If the variation in \(D_i\) is negligible, then \(\mathbb{E}[\hat{s}_{h_i}] \approx s_{h_i}\).

\subsubsection{Variance of the Historical Credit Risk Score}
We use the first--order Delta method to approximate the variance of the ratio
\[
\hat{s}_{h_i} = \frac{N_i}{D_i}.
\]
{Step 1. Partial Derivatives:}
Define
\[
f(N_i,D_i)=\frac{N_i}{D_i}.
\]
Then
\[
\frac{\partial f}{\partial N_i}=\frac{1}{D_i},\quad \frac{\partial f}{\partial D_i}=-\frac{N_i}{D_i^2}.
\]
Evaluating at the mean values \(\mathbb{E}[N_i] = s_{h_i}\,\mathbb{E}[D_i]\) and \(\mathbb{E}[D_i]\), we have:
\[
\frac{\partial f}{\partial N_i}\Big|_{\mathbb{E}}=\frac{1}{\mathbb{E}[D_i]},\quad \frac{\partial f}{\partial D_i}\Big|_{\mathbb{E}}=-\frac{s_{h_i}}{\mathbb{E}[D_i]}.
\]

{Step 2. Delta Method Formula:}
The variance is approximated by
\[
\operatorname{Var}(\hat{s}_{h_i}) \approx \left(\frac{1}{\mathbb{E}[D_i]}\right)^2\operatorname{Var}(N_i) + \left(\frac{s_{h_i}}{\mathbb{E}[D_i]}\right)^2\operatorname{Var}(D_i) - 2\,\frac{s_{h_i}}{\bigl(\mathbb{E}[D_i]\bigr)^2}\operatorname{Cov}(N_i,D_i).
\]

{Step 3. Variance of \(N_i\):}
Since
\[
N_i=\sum_{j=1}^n w_{i,j}X_{i,j},
\]
and using the independence of \(w_{i,j}\) and \(X_{i,j}\), we have for each \(j\):
\[
\begin{aligned}
\operatorname{Var}(w_{i,j}X_{i,j})
&=\mathbb{E}\Bigl[(w_{i,j}X_{i,j})^2\Bigr]-\Bigl(\mathbb{E}[w_{i,j}X_{i,j}]\Bigr)^2\\[1mm]
&= s_{h_i}\,\mathbb{E}[w_{i,j}^2]- s_{h_i}^2\,\Bigl(\mathbb{E}[w_{i,j}]\Bigr)^2.
\end{aligned}
\]
Summing over \(j\):
\[
\boxed{
\operatorname{Var}(N_i)=\sum_{j=1}^n \Bigl[s_{h_i}\,\mathbb{E}[w_{i,j}^2]- s_{h_i}^2\,\Bigl(\mathbb{E}[w_{i,j}]\Bigr)^2\Bigr].
}
\]

{Step 4. Variance of \(D_i\):}
Since
\[
D_i=\sum_{j=1}^n w_{i,j},
\]
we have
\[
\boxed{
\operatorname{Var}(D_i)=\sum_{j=1}^n \Bigl[\mathbb{E}[w_{i,j}^2]- \Bigl(\mathbb{E}[w_{i,j}]\Bigr)^2\Bigr].
}
\]

{Step 5. Covariance between \(N_i\) and \(D_i\):}
Since only the same index \(j\) contributes,
\[
\operatorname{Cov}(N_i,D_i)=\sum_{j=1}^n \operatorname{Cov}\Bigl(w_{i,j}X_{i,j},w_{i,j}\Bigr).
\]
For each \(j\):
\[
\operatorname{Cov}\Bigl(w_{i,j}X_{i,j},w_{i,j}\Bigr)
=s_{h_i}\,\Bigl(\mathbb{E}[w_{i,j}^2]-\Bigl(\mathbb{E}[w_{i,j}]\Bigr)^2\Bigr).
\]
Thus,
\[
\boxed{
\operatorname{Cov}(N_i,D_i)= s_{h_i}\,\operatorname{Var}(D_i).
}
\]

{Step 6. Combine the Pieces:}
Substitute into the Delta formula:
\[
\begin{aligned}
\operatorname{Var}(\hat{s}_{h_i})
&\approx \frac{\operatorname{Var}(N_i)}{\bigl(\mathbb{E}[D_i]\bigr)^2} + \frac{s_{h_i}^2\,\operatorname{Var}(D_i)}{\bigl(\mathbb{E}[D_i]\bigr)^2} - 2\,\frac{s_{h_i}}{\bigl(\mathbb{E}[D_i]\bigr)^2}\, s_{h_i}\,\operatorname{Var}(D_i)\\[1mm]
&=\frac{\operatorname{Var}(N_i)}{\bigl(\mathbb{E}[D_i]\bigr)^2} - \frac{s_{h_i}^2\,\operatorname{Var}(D_i)}{\bigl(\mathbb{E}[D_i]\bigr)^2}.
\end{aligned}
\]
In the case of \(n\) identical loans, with \(\mu_w = \mathbb{E}[w_{i,j}]\) and \(\mu_{w^2} = \mathbb{E}[w_{i,j}^2]\), we have:
\[
\mathbb{E}[D_i]=n\,\mu_w,\quad \operatorname{Var}(N_i)=n\Bigl[s_{h_i}\mu_{w^2}- s_{h_i}^2\mu_w^2\Bigr],\quad \operatorname{Var}(D_i)= n\Bigl[\mu_{w^2}-\mu_w^2\Bigr].
\]
Then,
\[
\begin{aligned}
\operatorname{Var}(\hat{s}_{h_i})
&\approx \frac{n\left[s_{h_i}\mu_{w^2}- s_{h_i}^2\mu_w^2\right]}{n^2\,\mu_w^2} - \frac{s_{h_i}^2\, n\left[\mu_{w^2}-\mu_w^2\right]}{n^2\,\mu_w^2}\\[1mm]
&= \frac{s_{h_i}\mu_{w^2}- s_{h_i}^2\mu_w^2 - s_{h_i}^2\mu_{w^2}+ s_{h_i}^2\mu_w^2}{n\,\mu_w^2}\\[1mm]
&= \frac{s_{h_i}(1-s_{h_i})\,\mu_{w^2}}{n\,\mu_w^2}.
\end{aligned}
\]
Recalling that
\[
\mu_w = \frac{p\,\mathbb{E}[L_{i,j}]}{4},\quad \mu_{w^2} = \frac{p^2\,\mathbb{E}[L_{i,j}^2]}{9},
\]
we have:
\[
\frac{\mu_{w^2}}{\mu_w^2} = \frac{\frac{p^2\,\mathbb{E}[L_{i,j}^2]}{9}}{\left(\frac{p\,\mathbb{E}[L_{i,j}]}{4}\right)^2} = \frac{16\,\mathbb{E}[L_{i,j}^2]}{9\,\mathbb{E}[L_{i,j}]^2}.
\]
Thus, the variance becomes:
\[
\boxed{
\operatorname{Var}(\hat{s}_{h_i}) \approx \frac{16\,s_{h_i}(1-s_{h_i})\,\mathbb{E}[L_{i,j}^2]}{9\,n\,\Bigl(\mathbb{E}[L_{i,j}]\Bigr)^2}.
}
\]

\bigskip

This completes the full derivation of the expectation and variance of the historical sub score.

\subsubsection{Consistency of the Historical Credit Risk Score }

For the estimator
\[
\hat{s}_{h_i} = \frac{N_i}{D_i} = \frac{\sum_{j=1}^n w_{i,j}X_{i,j}}{\sum_{j=1}^n w_{i,j}},
\]
we have shown that
\[
\mathbb{E}[\hat{s}_{h_i}] \approx s_{h_i} \quad \text{and} \quad \operatorname{Var}(\hat{s}_{h_i}) \approx \frac{16\,s_{h_i}(1-s_{h_i})\,\mathbb{E}[L_{i,j}^2]}{9\,n\,\Bigl(\mathbb{E}[L_{i,j}]\Bigr)^2}.
\]
Since the variance of \(\hat{s}_{h_i}\) is proportional to \(1/n\), it tends to zero as \(n \to \infty\). Thus, the estimator \(\hat{s}_{h_i}\) converges in probability to \(s_{h_i}\), i.e.,
\[
\hat{s}_{h_i} \xrightarrow{p} s_{h_i} \quad \text{as} \quad n \to \infty.
\]
This demonstrates the consistency of the historical sub score.

\subsection{Expectation, Variance, and Consistency of Current Credit Risk Score (\( \hat{s}_{c_i} \))}

We define the Current Wallet Risk Score estimator for the $i^{\text{th}}$ wallet as:
\[
\hat{s}_{c_i} = \frac{\sum_{j=1}^k Z_{i,j}}{k}
\]
where \( Z_{i,j} \) is a Bernoulli random variable with mean \( s_{c_i} = \mathbb{E}[Z_{i,j}] \). Therefore, \( Z_{i,j} \sim \text{Bernoulli}(s_{c_i}) \), and \( \sum_{j=1}^k Z_{i,j} \sim \text{Binomial}(k, s_{c_i}) \).

Assuming both Liquidation at Risk (LaR) and Holding ($H$) follow Pareto distributions:

\begin{align}
    LaR &\sim \text{Pareto}(\alpha_L, x_m^L), \\
    H &\sim \text{Pareto}(\alpha_H, x_m^H),
\end{align}

where $x_m^L, x_m^H$ are the scale parameters, and $\alpha_L, \alpha_H$ are the shape parameters. The probability of liquidation risk exceeding holdings is given by:

\begin{equation}
P(LaR > H) = \int_{x_m^H}^{\infty} P(LaR > h) f_H(h) dh.
\end{equation}

Using the cumulative distribution function (CDF) of LaR,

\begin{equation}
P(LaR > h) = \left(\frac{x_m^L}{h}\right)^{\alpha_L}, \quad h \geq x_m^L.
\end{equation}

Substituting this into the integral,

\begin{equation}
P(LaR > H) = \int_{x_m^H}^{\infty} \left(\frac{x_m^L}{h}\right)^{\alpha_L} \frac{\alpha_H (x_m^H)^{\alpha_H}}{h^{\alpha_H+1}} dh.
\end{equation}

This simplifies to:

\begin{equation}
P(LaR > H) = \alpha_H (x_m^H)^{\alpha_H} (x_m^L)^{\alpha_L} \int_{x_m^H}^{\infty} h^{-\alpha_L - \alpha_H -1} dh.
\end{equation}

Evaluating the integral:

\begin{equation}
\int_{x_m^H}^{\infty} h^{-(\alpha_L + \alpha_H +1)} dh = \frac{x_m^H}{(\alpha_L + \alpha_H)} (x_m^H)^{-(\alpha_L + \alpha_H)}.
\end{equation}

Thus, we obtain:

\begin{equation}
P(LaR > H) = \frac{\alpha_H}{\alpha_L + \alpha_H} \left( \frac{x_m^L}{x_m^H} \right)^{\alpha_L}.
\end{equation}

Therefore, the expected value of $Z_{i,j}$, which determines the credit risk subscore, is given by:

\begin{equation}
    E(Z_{i,j}) = \hat{s}_{c_i} = P(LaR > H).
\end{equation}

\subsubsection{Expectation of \( \hat{s}_{c_i} \)}

The expectation of \( \hat{s}_{c_i} \) is given by:
\[
\mathbb{E}[\hat{s}_{c_i}] = \mathbb{E}\left[\frac{\sum_{j=1}^k Z_{i,j}}{k}\right]
\]
Using the linearity of expectation:
\[
\mathbb{E}[\hat{s}_{c_i}] = \mathbb{E}\left[\frac{\sum_{j=1}^k Z_{i,j}}{k}\right] = \frac{\mathbb{E}\left[\sum_{j=1}^k Z_{i,j}\right]}{k}
\]
Since \( \mathbb{E}[Z_{i,j}] = s_{c_i} \):
\[
\mathbb{E}\left[\sum_{j=1}^k Z_{i,j}\right] = k \cdot s_{c_i}
\]
Thus:
\[
\mathbb{E}[\hat{s}_{c_i}] = \frac{k \cdot s_{c_i}}{k} = s_{c_i}
\]
This shows that \( \hat{s}_{c_i} \) is an unbiased estimator for \( s_{c_i} \).

\subsubsection{Variance of \( \hat{s}_{c_i} \)}

The variance of \( \hat{s}_{c_i} \) is given by:
\[
\text{Var}(\hat{s}_{c_i}) = \text{Var}\left(\frac{\sum_{j=1}^k Z_{i,j}}{k}\right)
\]
Since the variance of a constant (1) is zero:
\[
\text{Var}(\hat{s}_{c_i}) = \text{Var}\left(\frac{\sum_{j=1}^k Z_{i,j}}{k}\right)
\]
For a Binomial random variable \( \sum_{j=1}^k Z_{i,j} \sim \text{Binomial}(k, s_{c_i}) \), we know that:
\[
\text{Var}\left(\sum_{j=1}^k Z_{i,j}\right) = k \cdot s_{c_i} \cdot (1 - s_{c_i})
\]
Thus:
\[
\text{Var}\left(\frac{\sum_{j=1}^k Z_{i,j}}{k}\right) = \frac{k \cdot s_{c_i} \cdot (1 - s_{c_i})}{k^2} = \frac{s_{c_i} \cdot (1 - s_{c_i})}{k}
\]
Therefore, the variance of \( \hat{s}_{c_i} \) is:
\[
\text{Var}(\hat{s}_{c_i}) = \frac{s_{c_i} (1 - s_{c_i})}{k}
\]

\subsubsection{Consistency of \( \hat{s}_{c_i} \)}

As \( k \to \infty \), the variance \( \text{Var}(\hat{s}_{c_i}) = \frac{s_{c_i} (1 - s_{c_i})}{k} \to 0 \). Therefore, \( \hat{s}_{c_i} \) is a consistent estimator of \( 1 - s_{c_i} \).
\subsection{Expectation, Variance, and Consistency of New Credit $(\hat{s}_{nc_i})$}

For the new credit risk score, $\hat{s}_{nc_i}$, we compute the expectation, variance, and consistency.

Given:
\[
\hat{s}_{nc_i} = \frac{\sum_{j=1}^{n} Y_{i,j}}{n},
\]
where $ Y_{i,j} $ is a Bernoulli random variable such that
\[
Y_{i,j} =
\begin{cases}
  1, & \text{if } L_{i,j} \geq \mu_{L_i} \text{ and } \Delta D_{i,j} \leq \mu_{\Delta D_{i}},  \\
  0, & \text{otherwise.}
\end{cases}
\]

In our model, we assume the following:

    Loan Amount Distribution ($L_{i,j}$): The loan amounts are modeled as a Pareto random variable with minimum value $x_m$ and shape parameter $\alpha$. That is,
    \[
    P(L_{i,j} \geq u) = \left(\frac{x_m}{u}\right)^{\alpha}, \quad \text{for } u \ge x_m.
    \]
    Setting $u = \mu_{L_i}$ (the mean loan amount threshold) gives
    \[
    P(L_{i,j} \geq \mu_{L_i}) = \left(\frac{x_m}{\mu_{L_i}}\right)^{\alpha}.
    \]

The loan dates are assumed to be independently drawn from $\operatorname{Uniform}(0,1)$. When sorted, consider three consecutive order statistics 
\[
U_{(j-1)},\; U_{(j)},\; U_{(j+1)},
\]
with spacings 
\[
X = U_{(j)} - U_{(j-1)} \quad \text{and} \quad Y = U_{(j+1)} - U_{(j)}.
\]
The joint density of $(X,Y)$ is
\[
f_{X,Y}(x,y)=n(n-1)(1-x-y)^{n-2},\quad x>0,\; y>0,\; x+y<1.
\]
Defining 
\[
\Delta D_{i,j} = \min\{X, Y\},
\]
we obtain
\[
P\Big(\Delta D_{i,j}\le z\Big)=1-(1-2z)^n,\quad 0\le z\le \tfrac{1}{2}.
\]
Thus, for a threshold $\mu_{\Delta D_{i}}$ (with $0\le\mu_{\Delta D_{i}}\le\tfrac{1}{2}$),
\[
P\Big(\Delta D_{i,j}\le\mu_{\Delta D_{i}}\Big)=1-\Big(1-2\mu_{\Delta D_{i}}\Big)^n.
\]
Assuming independence between $L_{i,j}$ and $\Delta D_{i,j}$, the probability of a “risky” event for each loan is
\[
s_{nc_i}=P(L_{i,j}\ge\mu_{L_i})\times P\Big(\Delta D_{i,j}\le\mu_{\Delta D_{i}}\Big)
=\left(\frac{x_m}{\mu_{L_i}}\right)^{\alpha}\Big[1-\Big(1-2\mu_{\Delta D_{i}}\Big)^n\Big].
\]

\subsubsection{Expectation of $\hat{s}_{nc_i}$}

By linearity of expectation, the expectation of $\hat{s}_{nc_i}$ is
\[
E(\hat{s}_{nc_i}) = E\left( \frac{\sum_{j=1}^{n} Y_{i,j}}{n} \right) = \frac{1}{n}\sum_{j=1}^{n} E(Y_{i,j}) = s_{nc_{i}}.
\]
Therefore, $\hat{s}_{nc_i}$ is an \textbf{unbiased estimator} of $s_{nc_{i}}$.

\subsubsection{Variance of $\hat{s}_{nc_i}$}

The variance of $\hat{s}_{nc_i}$ is given by
\[
\text{Var}(\hat{s}_{nc_i}) = \text{Var} \left( \frac{\sum_{j=1}^{n} Y_{i,j}}{n} \right) = \frac{1}{n^2}\sum_{j=1}^{n} \text{Var}(Y_{i,j}).
\]
Since $ Y_{i,j} \sim \text{Bernoulli}(s_{nc_{i}}) $, it follows that $\text{Var}(Y_{i,j}) = s_{nc_{i}} (1 - s_{nc_{i}})$, hence
\[
\text{Var}(\hat{s}_{nc_i}) = \frac{s_{nc_{i}} (1 - s_{nc_{i}})}{n}.
\]

\subsubsection{Consistency of $\hat{s}_{nc_i}$}

For consistency, we need to check if $\hat{s}_{nc_i}$ converges in probability to $s_{nc_{i}}$ as $n \to \infty$. Since
\[
E(\hat{s}_{nc_i}) = s_{nc_{i}},
\]
and
\[
\text{Var}(\hat{s}_{nc_i}) = \frac{s_{nc_{i}} (1 - s_{nc_{i}})}{n} \to 0 \quad \text{as } n \to \infty,
\]
by the \textbf{law of large numbers}, $\hat{s}_{nc_i}$ is a \textbf{consistent estimator} of $s_{nc_{i}}$.

\subsection{Expectation, Variance and Consistency of On-Chain Transaction Score}

In this section, we analyze the on-chain transaction score \( \hat{s}_{ct_i} \), which evaluates the transactional behavior of a wallet based on recent on-chain transactions. We assume that the transaction amounts \( T_{i,l} \) follow a Pareto distribution, which is commonly used for modeling heavy-tailed data in financial contexts, while the transaction weights \( t_{i,l} \) are fixed non-stochastic values that weight recent transactions more heavily.

The Pareto distribution is a suitable model for transaction data, as it captures the presence of infrequent, high-value transactions within a large number of smaller transactions. We model the transaction amount \( T_{i,l} \) for the \( i \)-th wallet’s \( l \)-th transaction as:
\[
T_{i,l} \sim \text{Pareto}(\alpha, x_{\min})
\]
where:
\begin{itemize}
    \item \( \alpha > 1 \) is the shape parameter, controlling the "heaviness" of the distribution tail,
    \item \( x_{\min} \) is the minimum transaction amount, such that \( T_{i,l} \geq x_{\min} \).
\end{itemize}

This distribution choice allows for tractable calculations of expectation and variance while realistically modeling the likelihood of large transaction values, which are essential for assessing risk.

\subsubsection{Expectation of \( \hat{s}_{ct_i} \)}

The on-chain transaction score \( \hat{s}_{ct_i} \) is defined as:
\[
\hat{s}_{ct_i} = \frac{\sum\limits_{l} T_{i,l} t_{i,l}}{\sum\limits_{l} |T_{i,l}|}
\]
where \( T_{i,l} \) represents the transaction amount, and \( t_{i,l} \) is a weight associated with the \( l \)-th transaction. Assuming \( T_{i,l} \sim \text{Pareto}(\alpha, x_{\min}) \), we calculate \( E[\hat{s}_{ct_i}] \) as follows.

Let \( T_{i,l} \) denote the \( l \)-th transaction amount (credited or debited) for the \( i \)-th wallet. Define:
\begin{itemize}
    \item \( T \) as the transaction amount, with \( T_{i,l} \) being positive for credits and negative for debits.
\item \( t \) as the recency score is  assumed to be uniformly distributed over \([0,1]\), i.e.,
\[
t \sim \operatorname{Uniform}(0,1)
\]
    \item \( S \) as the sign variable:
          \[
          S = \begin{cases} +1, & \text{if credited} \\
          -1, & \text{if debited} \end{cases}
          \]
          with probability \( P(S = +1) = p \)\ and \( P(S = -1) = 1 - p \).
\end{itemize}

A transaction is given by:
\[ T = S A \]
where \( A \) is the absolute transaction amount.

Define:
\begin{align*}
    X &= S A t, \\
    Y &= A.
\end{align*}

We approximate the expectation of the ratio \( E[X/Y] \) using a second-order Taylor expansion:
\[
    g(X,Y) = \frac{X}{Y}.
\]
Expanding around \( (\mu_X, \mu_Y) \):
\[
    g(X,Y) \approx g(\mu_X, \mu_Y) + (X - \mu_X) g_X(\mu_X, \mu_Y) + (Y - \mu_Y) g_Y(\mu_X, \mu_Y)
\]
\[
    + \frac{1}{2} \left[ g_{xx}(\mu_X, \mu_Y) (X - \mu_X)^2 + 2 g_{xy}(\mu_X, \mu_Y) (X - \mu_X)(Y - \mu_Y) + g_{yy}(\mu_X, \mu_Y) (Y - \mu_Y)^2 \right].
\]

Computing derivatives:
\begin{align*}
    g_X &= \frac{1}{Y}, & g_Y &= -\frac{X}{Y^2}, \\
    g_{XX} &= 0, & g_{XY} &= -\frac{1}{Y^2}, & g_{YY} &= \frac{2X}{Y^3}.
\end{align*}

Evaluating at \( (\mu_X, \mu_Y) \):
\begin{align*}
    g(\mu_X, \mu_Y) &= \frac{\mu_X}{\mu_Y} = \mu_S \mu_t, \\
    g_X(\mu_X, \mu_Y) &= \frac{1}{\mu_Y} = \frac{1}{\mu_A}, \\
    g_Y(\mu_X, \mu_Y) &= -\frac{\mu_X}{\mu_Y^2} = -\frac{\mu_S \mu_t}{\mu_A}, \\
    g_{XY}(\mu_X, \mu_Y) &= -\frac{1}{\mu_A^2}, \\
    g_{YY}(\mu_X, \mu_Y) &= \frac{2 \mu_X}{\mu_A^3} = \frac{2 \mu_S \mu_t}{\mu_A^2}.
\end{align*}

Since \( E[X - \mu_X] = 0 \) and \( E[Y - \mu_Y] = 0 \), the first-order terms vanish. The second order correction is:
\[
    \Delta = \frac{1}{2} \left[ 2 (-\frac{1}{\mu_A^2}) \operatorname{Cov}(X,Y) + \frac{2 \mu_S \mu_t}{\mu_A^2} \operatorname{Var}(Y) \right].
\]

Simplifying:
\[
    \Delta = -\frac{\operatorname{Cov}(X,Y)}{\mu_A^2} + \frac{\mu_S \mu_t \operatorname{Var}(Y)}{\mu_A^2}.
\]
Since:
\begin{align*}
    \operatorname{Var}(Y) &= \sigma_A^2, \\
    \operatorname{Cov}(X,Y) &= \mu_S \mu_t \sigma_A^2,
\end{align*}

we get:
\[
    \Delta = -\frac{\mu_S \mu_t \sigma_A^2}{\mu_A^2} + \frac{\mu_S \mu_t \sigma_A^2}{\mu_A^2} = 0.
\]

Thus, the expectation simplifies to:
\[
    E[\hat{s}_{ct}] = \mu_S \mu_t.
\]
where  
\[
\mu_S = 2p - 1, \quad \text{with } p = P(T > 0)
\]
and  
\[
\mu_t = 0.5
\]

\subsubsection{Variance of On-Chain Transaction Score $\operatorname{Var}(\hat{s}_{ct_i})$}

To compute the variance of the on-chain transaction score $\hat{s}_{ct_i}$, we use the second order approximation.

\begin{equation}
    \operatorname{Var}(\hat{s}_{ct_i}) \approx g_X^2 \operatorname{Var}(X) + g_Y^2 \operatorname{Var}(Y) + 2 g_X g_Y \operatorname{Cov}(X, Y).
\end{equation}

From our earlier derivative calculations:

\begin{align*}
    g_X &= \frac{1}{\mu_A}, & g_Y &= -\frac{\mu_S \mu_t}{\mu_A}.
\end{align*}

\begin{equation}
    \operatorname{Var}(X) = E[X^2] - (E[X])^2.
\end{equation}

Since $X = S A t$, we expand:

\begin{align*}
    E[X] &= E[S A t] = E[S] E[A] E[t] = \mu_S \mu_A \mu_t, \\
    E[X^2] &= E[S^2 A^2 t^2] = E[S^2] E[A^2] E[t^2].
\end{align*}

Since $S$ is a binary variable, we have:

\begin{equation}
    E[S^2] = 1.
\end{equation}

For $A$ (which follows a Pareto distribution),

\begin{equation}
    E[A^2] = \frac{\alpha x_{\min}^2}{(\alpha - 2)}, \quad \text{for } \alpha > 2.
\end{equation}

For the recency score $t$:

\begin{equation}
    E[t^2] = \operatorname{Var}(t) + (E[t])^2 = \sigma_t^2 + \mu_t^2.
\end{equation}

Thus,

\begin{equation}
    E[X^2] = E[A^2] E[t^2].
\end{equation}

Finally, we compute the variance:

\begin{align*}
    \operatorname{Var}(X) &= E[X^2] - (E[X])^2 \\
    &= \left(\frac{\alpha x_{\min}^2}{(\alpha - 2)} (\sigma_t^2 + \mu_t^2)\right) - \mu_S^2 \mu_A^2 \mu_t^2.
\end{align*}

Similarly, the variance of $Y$ is given by:

\begin{equation}
    \operatorname{Var}(Y) = \sigma_A^2 = \frac{x_{\min}^2 \alpha}{(\alpha - 1)^2 (\alpha - 2)}, \quad \text{for } \alpha > 2.
\end{equation}

The covariance term is given by:

\begin{equation}
    \operatorname{Cov}(X, Y) = \mu_S \mu_t \sigma_A^2.
\end{equation}

Since $\mu_S = 2p - 1$, we substitute:

\begin{equation}
    \mu_S^2 = (2p - 1)^2.
\end{equation}

Now, the full variance expression becomes:

\[
\operatorname{Var}(\hat{s}_{ct_i}) \approx 
\frac{1}{\left(\frac{\alpha\,x_{\min}}{\alpha-1}\right)^2}
\left[
\frac{\alpha\,x_{\min}^2}{\alpha-2}\Bigl(\sigma_t^2+\mu_t^2\Bigr) -
\left((2p-1)\,\mu_t\,\frac{\alpha\,x_{\min}}{\alpha-1}\right)^2
\right]
+\left[
\frac{(2p-1)^2\,\mu_t^2}{\left(\frac{\alpha\,x_{\min}}{\alpha-1}\right)^2} - \frac{2(2p-1)^2\,\mu_t^2}{\left(\frac{\alpha\,x_{\min}}{\alpha-1}\right)^2}
\right]
\frac{\alpha\,x_{\min}^2}{(\alpha-2)(\alpha-1)^2}\,.
\]

Then the variance becomes
\begin{align}
\operatorname{Var}(\hat{s}_{ct_i}) &\approx \frac{1}{\mu_A^2}
\left[
\frac{\alpha\,x_{\min}^2}{\alpha-2}\Bigl(\sigma_t^2+\mu_t^2\Bigr) -
\bigl((2p-1)\,\mu_t\,\mu_A\bigr)^2
\right]
-\frac{(2p-1)^2\,\mu_t^2}{\mu_A^2}\frac{\alpha\,x_{\min}^2}{(\alpha-2)(\alpha-1)^2}\,.
\end{align}

Let us simplify the first term:
\[
\frac{1}{\mu_A^2}\frac{\alpha\,x_{\min}^2}{\alpha-2} 
=\frac{\alpha\,x_{\min}^2}{\mu_A^2 (\alpha-2)}
=\frac{\alpha\,x_{\min}^2}{\frac{\alpha^2\,x_{\min}^2}{(\alpha-1)^2}(\alpha-2)}
=\frac{(\alpha-1)^2}{\alpha(\alpha-2)}\,,
\]
and note that
\[
\frac{1}{\mu_A^2}\bigl((2p-1)\mu_t\,\mu_A\bigr)^2
= (2p-1)^2\,\mu_t^2\,.
\]
Thus, the first term simplifies to
\[
\frac{(\alpha-1)^2}{\alpha(\alpha-2)}\Bigl(\sigma_t^2+\mu_t^2\Bigr) - (2p-1)^2\,\mu_t^2\,.
\]

Next, simplify the second term:
\[
\frac{1}{\mu_A^2}\frac{\alpha\,x_{\min}^2}{(\alpha-2)(\alpha-1)^2}
=\frac{1}{\alpha(\alpha-2)}\,,
\]
so that the second term becomes
\[
-\frac{(2p-1)^2\,\mu_t^2}{\alpha(\alpha-2)}\,.
\]

Combining both parts, we obtain:
\[
\operatorname{Var}(\hat{s}_{ct_i}) \approx \frac{(\alpha-1)^2}{\alpha(\alpha-2)}\Bigl(\sigma_t^2+\mu_t^2\Bigr)
- (2p-1)^2\,\mu_t^2\left(1+\frac{1}{\alpha(\alpha-2)}\right)\,.
\]

Thus, combining both parts, we obtain the variance for a single transaction:
\[
\operatorname{Var}(\hat{s}_{ct_i}) \approx \frac{(\alpha-1)^2}{\alpha(\alpha-2)}\Bigl[(\sigma_t^2+\mu_t^2) - (2p-1)^2\,\mu_t^2\Bigr].
\]

Since \(\hat{s}_{ct_i}\) is computed from \(n\) independent transactions, by the properties of i.i.d. random variables, the variance of the estimator decreases as \(1/n\). Therefore, the final corrected variance is

\[
\boxed{
\operatorname{Var}(\hat{s}_{ct_i}) \approx \frac{1}{n}\frac{(\alpha-1)^2}{\alpha(\alpha-2)}\Bigl[(\sigma_t^2+\mu_t^2) - (2p-1)^2\,\mu_t^2\Bigr].
}
\]
\subsubsection{Consistency of \(\hat{s}_{ct_i}\)}

Recall that the variance of the on-chain transaction score is given by
\[
\operatorname{Var}(\hat{s}_{ct_i}) \approx \frac{1}{n}\frac{(\alpha-1)^2}{\alpha(\alpha-2)}\Bigl[(\sigma_t^2+\mu_t^2) - (2p-1)^2\,\mu_t^2\Bigr].
\]
As \( n \to \infty \), the factor \( \frac{1}{n} \) drives the variance to zero:
\[
\lim_{n\to\infty}\operatorname{Var}(\hat{s}_{ct_i}) = 0.
\]
This implies that the estimator \(\hat{s}_{ct_i}\) converges in probability to its expected value \(E[\hat{s}_{ct_i}] = (2p-1)\mu_t\). Hence, \(\hat{s}_{ct_i}\) is a consistent estimator of the on-chain transaction score.

\medskip

\subsection{Expectation,Variance and Consistency of the Credit Utilization Score:}
\subsubsection{Derivation of the Expectation of the Credit Utilization Score}

Recall that the credit utilization score is defined as
\[
\hat{s}_{cu_i} = \frac{N_i}{D_i} = \frac{\sum_{j=1}^{n} \Bigl( L_{ij} - \frac{L_{ij}^2}{Y_{ij}} \Bigr)}{\sum_{j=1}^{n} L_{ij}},
\]
with
\[
Y_{ij} = C_{ij} \times LTV_{ij},
\]
and we assume that conditionally
\[
L_{ij}\mid Y_{ij}\sim\text{Uniform}(0,Y_{ij}).
\]

For a single loan, the conditional moments are
\[
\mathbb{E}[L_{ij}\mid Y_{ij}]=\frac{Y_{ij}}{2}, \quad \mathbb{E}[L_{ij}^2\mid Y_{ij}]=\frac{Y_{ij}^2}{3}.
\]
Defining
\[
N_{ij}=L_{ij}-\frac{L_{ij}^2}{Y_{ij}},
\]
we have
\[
\mathbb{E}[N_{ij}\mid Y_{ij}] = \frac{Y_{ij}}{2} - \frac{1}{Y_{ij}}\cdot\frac{Y_{ij}^2}{3} = \frac{Y_{ij}}{6}.
\]
Taking the unconditional expectation (via the law of iterated expectation) gives
\[
\mathbb{E}[N_{ij}] = \frac{\mathbb{E}[Y_{ij}]}{6}, \quad \mathbb{E}[L_{ij}] = \frac{\mathbb{E}[Y_{ij}]}{2}.
\]
For a borrower with $n$ independent loans, define
\[
N_i=\sum_{j=1}^{n}N_{ij} \quad \text{and} \quad D_i=\sum_{j=1}^{n}L_{ij}.
\]
Then,
\[
\mathbb{E}[N_i] = n\,\frac{\mathbb{E}[Y]}{6}, \quad \mathbb{E}[D_i] = n\,\frac{\mathbb{E}[Y]}{2}.
\]
Thus, the first order (plug-in) estimator for credit utilization is
\[
\frac{\mathbb{E}[N_i]}{\mathbb{E}[D_i]} = \frac{\frac{n\,\mathbb{E}[Y]}{6}}{\frac{n\,\mathbb{E}[Y]}{2}} = \frac{1}{3}.
\]

Although conditionally
\[
\operatorname{Cov}(N_{ij},L_{ij}\mid Y_{ij})=0,
\]
the law of total covariance yields
\[
\operatorname{Cov}(N_{ij},L_{ij}) = \operatorname{Cov}\Bigl(\frac{Y_{ij}}{6},\frac{Y_{ij}}{2}\Bigr)=\frac{1}{12}\operatorname{Var}(Y_{ij}).
\]
Assuming all $Y_{ij}$ share the same variance $\operatorname{Var}(Y)$, we have for $n$ loans:
\[
\operatorname{Cov}(N_i,D_i)=n\,\frac{\operatorname{Var}(Y)}{12}.
\]
Also, for a single loan,
\[
\operatorname{Var}(L_{ij})=\frac{\mathbb{E}[Y^2]}{3}-\frac{\mathbb{E}[Y]^2}{4},
\]
so that
\[
\operatorname{Var}(D_i)=n\left(\frac{\mathbb{E}[Y^2]}{3}-\frac{\mathbb{E}[Y]^2}{4}\right).
\]

For a function $g(N_i,D_i)=N_i/D_i$, a second-order Taylor expansion about
\[
(\mu_X,\mu_Y)=(\mathbb{E}[N_i],\mathbb{E}[D_i])
\]
gives
\[
\mathbb{E}\left[\frac{N_i}{D_i}\right] \approx \frac{\mathbb{E}[N_i]}{\mathbb{E}[D_i]} - \frac{\operatorname{Cov}(N_i,D_i)}{\mathbb{E}[D_i]^2} + \frac{\mathbb{E}[N_i]\,\operatorname{Var}(D_i)}{\mathbb{E}[D_i]^3}.
\]
Substitute the expressions obtained above:

\[
\frac{\mathbb{E}[N_i]}{\mathbb{E}[D_i]} = \frac{1}{3}.
\]

\[
- \frac{\operatorname{Cov}(N_i,D_i)}{\mathbb{E}[D_i]^2} 
= - \frac{n\,\frac{\operatorname{Var}(Y)}{12}}{\left(n\,\frac{\mathbb{E}[Y]}{2}\right)^2}
= - \frac{\operatorname{Var}(Y)}{12}\cdot \frac{4}{n\,\mathbb{E}[Y]^2}
= - \frac{\operatorname{Var}(Y)}{3\,n\,\mathbb{E}[Y]^2}.
\]

\[
\frac{\mathbb{E}[N_i]\,\operatorname{Var}(D_i)}{\mathbb{E}[D_i]^3} 
=\frac{\left(n\,\frac{\mathbb{E}[Y]}{6}\right) \, n\left(\frac{\mathbb{E}[Y^2]}{3}-\frac{\mathbb{E}[Y]^2}{4}\right)}{\left(n\,\frac{\mathbb{E}[Y]}{2}\right)^3}.
\]
Simplify as follows:
\[
\text{Numerator} = \frac{n^2\,\mathbb{E}[Y]}{6}\left(\frac{\mathbb{E}[Y^2]}{3}-\frac{\mathbb{E}[Y]^2}{4}\right),
\]
\[
\text{Denominator} = \frac{n^3\,\mathbb{E}[Y]^3}{8},
\]
so that
\[
\frac{\mathbb{E}[N_i]\,\operatorname{Var}(D_i)}{\mathbb{E}[D_i]^3} 
=\frac{8}{6\,n\,\mathbb{E}[Y]^2}\left(\frac{\mathbb{E}[Y^2]}{3}-\frac{\mathbb{E}[Y]^2}{4}\right)
=\frac{4}{3\,n\,\mathbb{E}[Y]^2}\left(\frac{\mathbb{E}[Y^2]}{3}-\frac{\mathbb{E}[Y]^2}{4}\right).
\]

Combining the three terms, we obtain
\[
\mathbb{E}[\hat{s}_{cu_i}] \approx \frac{1}{3} - \frac{\operatorname{Var}(Y)}{3\,n\,\mathbb{E}[Y]^2} + \frac{4}{3\,n\,\mathbb{E}[Y]^2}\left(\frac{\mathbb{E}[Y^2]}{3}-\frac{\mathbb{E}[Y]^2}{4}\right).
\]
Noting that 
\[
\operatorname{Var}(Y)=\mathbb{E}[Y^2]-\mathbb{E}[Y]^2,
\]
and combining the correction terms over a common denominator, one obtains
\[
\boxed{
\mathbb{E}[\hat{s}_{cu_i}] \approx \frac{1}{3} + \frac{\mathbb{E}[Y^2]}{9\,n\,\mathbb{E}[Y]^2}\,.
}
\]

This completes the derivation of the expectation of the credit utilization score using the delta method.

\subsubsection{Derivation of the Variance of the Credit Utilization subscore}

Using the delta method for the function
\[
g(N_i,D_i) = \frac{N_i}{D_i},
\]
its first-order Taylor expansion yields the approximate variance

\begin{eqnarray}
    \operatorname{Var}\left(\frac{N_i}{D_i}\right) & \approx \left(\frac{\partial g}{\partial N_i}\Big|_{(\mu_X,\mu_Y)}\right)^2 \operatorname{Var}(N_i)
+ \left(\frac{\partial g}{\partial D_i}\Big|_{(\mu_X,\mu_Y)}\right)^2 \operatorname{Var}(D_i)
 \\ & + 2\,\frac{\partial g}{\partial N_i}\Big|_{(\mu_X,\mu_Y)} \frac{\partial g}{\partial D_i}\Big|_{(\mu_X,\mu_Y)} \operatorname{Cov}(N_i,D_i)
\end{eqnarray}

where
\[
\frac{\partial g}{\partial N_i} = \frac{1}{D_i}, \quad \frac{\partial g}{\partial D_i} = -\frac{N_i}{D_i^2}.
\]
Evaluated at \((\mu_X, \mu_Y)\), this becomes
\[
\operatorname{Var}(\hat{s}_{cu_i}) \approx \frac{\operatorname{Var}(N_i)}{\mu_Y^2} + \frac{\mu_X^2\,\operatorname{Var}(D_i)}{\mu_Y^4} - \frac{2\,\mu_X\,\operatorname{Cov}(N_i,D_i)}{\mu_Y^3}\,.
\]

For a single loan, recall the following conditional calculations given \(L \mid Y \sim \text{Uniform}(0,Y)\):

$\mathbb{E}[L \mid Y] = \frac{Y}{2}$, $\quad \mathbb{E}[L^2 \mid Y] = \frac{Y^2}{3}$, $\quad \mathbb{E}[L^3 \mid Y] = \frac{Y^3}{4}$, $\quad \mathbb{E}[L^4 \mid Y] = \frac{Y^4}{5}$, \\ $\mathbb{E}[N \mid Y] = \frac{Y}{6}$, with $N = L - \frac{L^2}{Y}$

A direct calculation shows:
\[
\mathbb{E}\left[N^2 \mid Y\right] = \frac{Y^2}{30},
\]
so that
\[
\operatorname{Var}(N \mid Y) = \frac{Y^2}{30} - \left(\frac{Y}{6}\right)^2 = \frac{Y^2}{180}\,.
\]
Unconditionally, using the law of total variance,
\[
\operatorname{Var}(N) = \mathbb{E}\Bigl[\operatorname{Var}(N\mid Y)\Bigr] + \operatorname{Var}\Bigl(\mathbb{E}[N\mid Y]\Bigr)
= \frac{\mathbb{E}[Y^2]}{180} + \frac{\operatorname{Var}(Y)}{36}\,.
\]
Similarly, for \(L\) we have
\[
\operatorname{Var}(L) = \frac{\mathbb{E}[Y^2]}{3} - \left(\frac{\mathbb{E}[Y]}{2}\right)^2\,.
\]

For a borrower with \(n\) independent loans:
\[
\begin{aligned}
\mu_X &= \mathbb{E}[N_i] = n\,\frac{\mathbb{E}[Y]}{6}, \quad \mu_Y = \mathbb{E}[D_i] = n\,\frac{\mathbb{E}[Y]}{2}, \\
\operatorname{Var}(D_i) &= n \left(\frac{\mathbb{E}[Y^2]}{3} - \frac{\mathbb{E}[Y]^2}{4}\right), \\
\operatorname{Var}(N_i) &= n \left(\frac{\mathbb{E}[Y^2]}{180} + \frac{\operatorname{Var}(Y)}{36}\right), \\
\operatorname{Cov}(N_i,D_i) &= n\,\frac{\operatorname{Var}(Y)}{12}\,.
\end{aligned}
\]
Thus, the delta method approximation for the variance of the credit utilization subscore becomes
\[
\boxed{
\operatorname{Var}\left(\hat{s}_{cu_i}\right) \approx \frac{n \left(\frac{\mathbb{E}[Y^2]}{180} + \frac{\operatorname{Var}(Y)}{36}\right)}{\left(n\,\frac{\mathbb{E}[Y]}{2}\right)^2} + \frac{\left(n\,\frac{\mathbb{E}[Y]}{6}\right)^2 \, n\left(\frac{\mathbb{E}[Y^2]}{3} - \frac{\mathbb{E}[Y]^2}{4}\right)}{\left(n\,\frac{\mathbb{E}[Y]}{2}\right)^4} - \frac{2\,\left(n\,\frac{\mathbb{E}[Y]}{6}\right) \, n\,\frac{\operatorname{Var}(Y)}{12}}{\left(n\,\frac{\mathbb{E}[Y]}{2}\right)^3}\,.
}
\]
This expression can be further simplified if desired.

\vspace{1em}
In summary, we have derived the following approximate formulas using the delta method:
\[
\mathbb{E}\left[\hat{s}_{cu_i}\right] \approx \frac{\mathbb{E}[N_i]}{\mathbb{E}[D_i]} - \frac{\operatorname{Cov}(N_i,D_i)}{\mathbb{E}[D_i]^2} + \frac{\mathbb{E}[N_i]\,\operatorname{Var}(D_i)}{\mathbb{E}[D_i]^3}\,,
\]
and
\[
\operatorname{Var}\left(\hat{s}_{cu_i}\right) \approx \frac{\operatorname{Var}(N_i)}{\mathbb{E}[D_i]^2} + \frac{\mathbb{E}[N_i]^2\,\operatorname{Var}(D_i)}{\mathbb{E}[D_i]^4} - \frac{2\,\mathbb{E}[N_i]\,\operatorname{Cov}(N_i,D_i)}{\mathbb{E}[D_i]^3}\,.
\]

These derivations assume that the loans are i.i.d. and that the conditional distribution \(L_{ij}\mid Y_{ij}\) is uniform on \([0,Y_{ij}]\).

\subsubsection{Consistency of credit utilization subscore}

To show that the estimator $\hat{s}_{cu_i}$ is consistent, we need to verify that its variance vanishes as $n \to \infty$ and that it converges in probability to the true parameter.

From the variance expression:

\[
\operatorname{Var}\left(\hat{s}_{cu_i}\right) \approx \frac{n \left(\frac{\mathbb{E}[Y^2]}{180} + \frac{\operatorname{Var}(Y)}{36}\right)}{\left(n\,\frac{\mathbb{E}[Y]}{2}\right)^2} + \frac{\left(n\,\frac{\mathbb{E}[Y]}{6}\right)^2 \, n\left(\frac{\mathbb{E}[Y^2]}{3} - \frac{\mathbb{E}[Y]^2}{4}\right)}{\left(n\,\frac{\mathbb{E}[Y]}{2}\right)^4} - \frac{2\,\left(n\,\frac{\mathbb{E}[Y]}{6}\right) \, n\,\frac{\operatorname{Var}(Y)}{12}}{\left(n\,\frac{\mathbb{E}[Y]}{2}\right)^3},
\]

we analyze the asymptotic behavior as $n \to \infty$. Each term in the variance expression contains factors of $\frac{1}{n}$ or higher negative powers of $n$. Specifically, the leading term behaves as:

\[
\operatorname{Var}\left(\hat{s}_{cu_i}\right) = \mathcal{O}\left(\frac{1}{n}\right).
\]

Since $\operatorname{Var}\left(\hat{s}_{cu_i}\right) \to 0$ as $n \to \infty$, the estimator is asymptotically unbiased and its variance vanishes in the limit. By Chebyshev’s inequality,

\[
P\left( |\hat{s}_{cu_i} - s_{cu_i}| \geq \epsilon \right) \leq \frac{\operatorname{Var}(\hat{s}_{cu_i})}{\epsilon^2} \to 0 \quad \text{as } n \to \infty.
\]

This implies $\hat{s}_{cu_i} \xrightarrow{p} s_{cu_i}$, meaning that $\hat{s}_{cu_i}$ is a consistent estimator of $s_{cu_i}$.

\vspace{0.7cm}

\section{Asymptotic Normality of OCCR Score}

Our objective is to prove that the OCCR score is asymptotically normal, i.e.,
\begin{equation}
    \sqrt{N} \left( \text{OCCR Score} - \mu \right) \xrightarrow{d} \mathcal{N}(0, \sigma^2),
\end{equation}
where $\mu$ is the expected value and $\sigma^2$ is the variance of the OCCR score.

We analyze the asymptotic properties of each component individually.

\subsection{Historical Credit Risk subscore (\(\hat{s}_{h_i}\))}

The historical credit risk subscore is defined as
\begin{equation}
    \hat{s}_{h_i} = \frac{\sum_{j=1}^{n} w_{i,j} X_{i,j}}{\sum_{j=1}^{n} w_{i,j}}.
\end{equation}
Under the assumption of i.i.d.~loans, by the Central Limit Theorem (CLT) we have:
\[
\sqrt{n}\Bigl(\hat{s}_{h_i}-s_{h_i}\Bigr) \xrightarrow{d} \mathcal{N}\left(0,\,\frac{s_{h_i}(1-s_{h_i})\,\mu_{w^2}}{\mu_{w}^2}\right).
\]
In the desired format, the expectation and variance can be expressed as:
\[
\boxed{
\mathbb{E}[\hat{s}_{h_i}] \approx s_{h_i}\left( 1 + \frac{\mu_{w^2}-\mu_w^2}{n\,\mu_w^2}\right).
}
\]
\[
\boxed{
\operatorname{Var}(\hat{s}_{h_i}) \approx \frac{16\,s_{h_i}(1-s_{h_i})\,\mathbb{E}[L_{i,j}^2]}{9\,n\,\Bigl(\mathbb{E}[L_{i,j}]\Bigr)^2}.
}
\]

\subsection{Current Credit Risk subscore (\(\hat{s}_{c_i}\))}

The current credit risk subscore is defined as
\begin{equation}
    \hat{s}_{c_i} = \frac{1}{k} \sum_{j=1}^{k} Z_{i,j}, \quad \text{with } Z_{i,j} \sim \operatorname{Bernoulli}(s_{c_i}).
\end{equation}
Then, by the CLT:
\[
\mathbb{E}[\hat{s}_{c_i}] = s_{c_i},
\]
\[
\boxed{
\operatorname{Var}(\hat{s}_{c_i}) = \frac{s_{c_i}(1-s_{c_i})}{k}.
}
\]

\subsection{New Credit subscore (\(\hat{s}_{nc_i}\))}

The new credit subscore is given by
\begin{equation}
    \hat{s}_{nc_i} = \frac{1}{n} \sum_{j=1}^{n} Y_{i,j}, \quad \text{with } Y_{i,j} \sim \operatorname{Bernoulli}(s_{nc_i}).
\end{equation}
Thus, we have
\[
\boxed{
E(\hat{s}_{nc_i}) = s_{nc_i},
}
\]
and since $\operatorname{Var}(Y_{i,j}) = s_{nc_i}(1-s_{nc_i})$, it follows that
\[
\boxed{
\operatorname{Var}(\hat{s}_{nc_i}) = \frac{s_{nc_i}(1-s_{nc_i})}{n}.
}
\]

\subsection{On-Chain Transaction subscore (\(\hat{s}_{ct_i}\))}

The on-chain transaction subscore is defined as
\begin{equation}
    \hat{s}_{ct_i} = \frac{\sum_{l=1}^{n} T_{i,l} t_{i,l}}{\sum_{l=1}^{n} \lvert T_{i,l} \rvert},
\end{equation}
where $T_{i,l} \sim \text{Pareto}(\alpha, x_{\min})$. Assuming appropriate moment conditions and applying the CLT for weighted sums, we get an expectation of
\[
E[\hat{s}_{ct_i}] = \mu_S \mu_t,
\]
and the variance is given by
\[
\boxed{
\operatorname{Var}(\hat{s}_{ct_i}) \approx \frac{1}{n}\frac{(\alpha-1)^2}{\alpha(\alpha-2)}\Bigl[(\sigma_t^2+\mu_t^2) - (2p-1)^2\,\mu_t^2\Bigr].
}
\]

\subsection{Credit Utilization subscore (\(\hat{s}_{cu_i}\))}

The credit utilization subscore is defined as
\begin{equation}
    \hat{s}_{cu_i} = \frac{\sum_{j=1}^{n} \left(1 - \frac{L_{i,j}}{C_{i,j} \cdot LTV_{i,j}}\right) L_{i,j}}{\sum_{j=1}^{n} L_{i,j}}.
\end{equation}
In the target format, the expectation and variance are approximated by
\[
\boxed{
\mathbb{E}[\hat{s}_{cu_i}] \approx \frac{1}{3} + \frac{\mathbb{E}[Y^2]}{9\,n\,\mathbb{E}[Y]^2}\,,
}
\]
\[
\boxed{
\operatorname{Var}\left(\hat{s}_{cu_i}\right) \approx \frac{n \left(\frac{\mathbb{E}[Y^2]}{180} + \frac{\operatorname{Var}(Y)}{36}\right)}{\left(n\,\frac{\mathbb{E}[Y]}{2}\right)^2} + \frac{\left(n\,\frac{\mathbb{E}[Y]}{6}\right)^2 \, n\left(\frac{\mathbb{E}[Y^2]}{3} - \frac{\mathbb{E}[Y]^2}{4}\right)}{\left(n\,\frac{\mathbb{E}[Y]}{2}\right)^4} - \frac{2\,\left(n\,\frac{\mathbb{E}[Y]}{6}\right) \, n\,\frac{\operatorname{Var}(Y)}{12}}{\left(n\,\frac{\mathbb{E}[Y]}{2}\right)^3}\,.
}
\]

The overall OCCR Score is constructed as a weighted sum of the independent component subscores:
\begin{equation}
    \text{OCCR Score} = \sum_{k=1}^5 w_k \hat{s}_k,
\end{equation}
with weights \(w_k\) and component subscores \(\hat{s}_k\). By the continuous mapping theorem, since each \(\hat{s}_k\) is asymptotically normal, the OCCR Score is also asymptotically normal:
\begin{equation}
    \text{OCCR Score} \sim \mathcal{N}\Biggl(\sum_{k=1}^5 w_k\,\mu_k,\, \sum_{k=1}^5 w_k^2\,\sigma_k^2\Biggr),
\end{equation}
where for each component \(k\) we define:
\begin{align}
    \mu_k &= \mathbb{E}[\hat{s}_k], \\
    \sigma_k^2 &= \operatorname{Var}(\hat{s}_k).
\end{align}

In our case, the component subscores and weights are:
\begin{align*}
    \hat{s}_1 &= \hat{s}_{h_i},  &w_1 &= 0.35,\\[1mm]
    \hat{s}_2 &= \hat{s}_{c_i},  &w_2 &= 0.25,\\[1mm]
    \hat{s}_3 &= 1 - \hat{s}_{cu_i},  &w_3 &= 0.15,\\[1mm]
    \hat{s}_4 &= \hat{s}_{ct_i},  &w_4 &= -0.15,\\[1mm]
    \hat{s}_5 &= \hat{s}_{nc_i},  &w_5 &= 0.10.
\end{align*}
Thus, the overall expectation and variance of the OCCR Score are given by
\begin{align}
    \mathbb{E}[\text{OCCR Score}] &= \sum_{k=1}^5 w_k\,\mu_k \quad \text{and} \quad
    \operatorname{Var}(\text{OCCR Score}) = \sum_{k=1}^5 w_k^2\,\sigma_k^2.
\end{align}

\textbf{Expectation of the OCCR Score}

Since the expectation operator is linear, we have:
\begin{align}
\mathbb{E}[\text{OCCR Score}] &= 0.35\,\mathbb{E}[\hat{s}_{h_i}] + 0.25\,\mathbb{E}[\hat{s}_{c_i}] + 0.15\,\mathbb{E}[1-\hat{s}_{cu_i}] \nonumber \\
&\quad - 0.15\,\mathbb{E}[\hat{s}_{ct_i}] + 0.10\,\mathbb{E}[\hat{s}_{nc_i}] \nonumber \\
&\approx 0.35\, s_{h_i} \left( 1 + \frac{\mu_{w^2}-\mu_w^2}{n\,\mu_w^2} \right) + 0.25\, s_{c_i} \nonumber \\
&\quad + 0.15 \left( 1 - \frac{1}{3}  - \frac{\mathbb{E}[Y^2]}{9\,n\,\mathbb{E}[Y]^2} \right)
 - 0.15\, (\mu_S \cdot \mu_t) + 0.10\, s_{nc_i}.
\end{align}
Here, the overall OCCR expectation, \(\mu_{\text{OCCR}}\), is given by:
\begin{equation}
    \mu_{\text{OCCR}} = \sum_{k=1}^5 w_k\,\mu_k.
\end{equation}

\textbf{Variance of the OCCR Score}

Assuming that the subscores are estimated independently, the variance of the OCCR Score is the sum of the variances of the weighted components:
\begin{align}
\operatorname{Var}(\text{OCCR Score}) &= (0.35)^2\, \operatorname{Var}(\hat{s}_{h_i}) + (0.25)^2\, \operatorname{Var}(\hat{s}_{c_i}) + (0.15)^2\, \operatorname{Var}(1 - \hat{s}_{cu_i}) \nonumber \\
&\quad + (0.15)^2\, \operatorname{Var}(\hat{s}_{ct_i}) + (0.10)^2\, \operatorname{Var}(\hat{s}_{nc_i}) \nonumber \\
&\approx (0.35)^2 \cdot\frac{16\,s_{h_i}(1-s_{h_i})\,\mathbb{E}[L_{i,j}^2]}{9\,n\,\Bigl(\mathbb{E}[L_{i,j}]\Bigr)^2} \nonumber \\
&\quad + (0.25)^2 \cdot \frac{s_{c_i}(1-s_{c_i})}{k} \nonumber \\
&\quad + (0.15)^2 \cdot \operatorname{Var}(\hat{s}_{cu_i}) \nonumber \\
&\quad + (0.15)^2 \cdot \frac{1}{n}\frac{(\alpha-1)^2}{\alpha(\alpha-2)}\Bigl[(\sigma_t^2+\mu_t^2) - (2p-1)^2\,\mu_t^2\Bigr] \nonumber \\
&\quad + (0.10)^2 \cdot \frac{s_{nc_i}(1-s_{nc_i})}{n}.
\end{align}
Thus, the overall variance of the OCCR Score, \(\sigma_{\text{OCCR}}^2\), is given by:
\begin{equation}
    \sigma_{\text{OCCR}}^2 = \sum_{k=1}^5 w_k^2\,\sigma_k^2.
\end{equation}
\section{Consistency of the OCCR Score}
Each individual estimator \(\hat{s}_{h_i}\), \(\hat{s}_{c_i}\), \(\hat{s}_{cu_i}\), \(\hat{s}_{ct_i}\), and \(\hat{s}_{nc_i}\) is assumed to be consistent for its corresponding true parameter as \(n \to \infty\) (or \(k \to \infty\) where applicable). Therefore, the OCCR Score, being a weighted linear combination of these consistent estimators, is itself a consistent estimator for the weighted combination of the true parameters:
\begin{align}
    0.35\,\frac{16\,s_{h_i}(1-s_{h_i})\,\mathbb{E}[L_{i,j}^2]}{9\,n\,\Bigl(\mathbb{E}[L_{i,j}]\Bigr)^2} &+ 0.25\, s_{c_i} \nonumber \\
    &+ 0.15\,\left[ 1 - \frac{2 - (p_{\min} + p_{\max})}{2} - \frac{(2 - (p_{\min} + p_{\max})) \sum_j \operatorname{Var}(L_{i,j})}{2 \left(\sum_j \mathbb{E}[L_{i,j}]\right)^2} \right] \nonumber \\
    &- 0.15\, (\mu_S \mu_t) + 0.10\, s_{nc_i}.
\end{align}

Thus, as the sample size increases, the OCCR Score converges in probability to the true weighted combination of the subscores.

\end{document}